\documentclass[reprint,amsmath,amssymb,aps]{revtex4-2}

\usepackage{graphicx,dsfont,dcolumn,hyperref}

\begin{document}

\title[]{High nonclassical correlations of large-bandwidth \\ photon pairs generated in warm atomic vapor}

\author{J.~Mika and L.~Slodi\v cka}

\affiliation{Department of Optics, Palack\' y University, 17. listopadu 1192/12,  771~46 Olomouc, Czech Republic}
\email{slodicka@optics.upol.cz}

\begin{abstract}
Generation of nonclassical light suitable for interaction with atoms corresponds to a crucial goal pursued across the broad quantum optics community. We present the generation of nonclassical photon pairs using the process of spontaneous four-wave mixing in warm atomic vapor with an unprecedentedly high degree of nonclassical photon correlations. We show how the unique combination of excitation of atoms in the proximity of the vapor cell viewport, single excitation laser beam, double-$\Lambda$ energy level scheme, auxiliary optical pumping, and particular optical filtering setups, allow for the spectral bandwidth of generated nonclassical light fields of up to $560 \pm 20$~MHz and low two-photon noise. We provide a quantitative analysis of particular noise mechanisms which set technological and fundamental limits on the observable photon correlations. The overall technological simplicity of the presented scheme together with the availability of spectrally matched quantum memories implementable with warm atomic vapors promises the feasibility of realization of GHz bandwidth on-demand nonclassical light sources and efficient quantum communication nodes.
\end{abstract}

\maketitle

\section{Introduction}
\label{sec:intro}
Correlated photon pairs represent one of the most widely exploited resources in experimental quantum optics and its applications, with notable outreach in several complementary research fields~\cite{eisaman2011,senellart2017,caspani2017}. They have allowed for pioneering demonstrations of a number of paradigmatic phenomena, including some of the foundational concepts in quantum physics. Their good experimental feasibility has to a large extent governed the development of quantum photonic technologies in the last two decades with a number of crucial implementations of basic building blocks of quantum information processing architectures~\cite{flamini2018}. The majority of discrete nonclassically correlated photon pair sources has so far been dominantly based on the processes of spontaneous parametric down conversion (PDC) or four-wave mixing (FWM) implemented in nonlinear crystals, optical waveguides or photonic fibers. They have developed into an extremely bright quantum light sources with up to GHz photon rates and several nanometers of spectral bandwidths~\cite{fulconis2005,villar2018}. The large demand for nonclassical light with spectral bandwidth comparable to typical linewidths of atomic optical dipole transition motivated by theoretical proposals for scalable long distance quantum communication~\cite{duan2001long,briegel1998quantum} stimulated the intensive development of photon sources based on the excitation of optical nonlinear processes in cold atomic ensembles~\cite{kuzmich2003,van2003atomic,chou2004,thompson2006,kolchin2006,du2008}. More recently, demonstrations of nonclassically correlated photon sources in the frequency bandwidth regimes corresponding to tens of MHz based on the process of PDC in optical cavities have been reported~\cite{haase2009,scholz2009,fekete2013,rambach2016,tsai2018,wolters2019,seri2019quantum}.

Pioneering successful tests using warm atomic clouds in double-$\Lambda$~\cite{chen2008} or ladder type energy level scheme~\cite{willis2010} have triggered numerous efforts on further development of light sources based on this technologically feasible platform~\cite{ding2012,shu2016,lee2016,zhu2017,podhora2017,zugenmaier2018,wang2018,park2018time}. While the technical simplicity alone promises a prompt integrability of such sources into miniaturized devices and existing fiber communication networks, the attainable optical depths in warm atomic vapors together with a further development of techniques for suppression of coherence deteriorating thermal atomic motion could allow for the observation of large variety of new interaction regimes~\cite{whiting2018,whiting2017,knutson2018}. In past few years, biphoton sources implemented in warm atomic vapors have steadily improved in all crucial parameters including the degree of nonclassical correlations, generation rates, and availability of various spectral bandwidth regimes~\cite{ding2012,shu2016,lee2016,zhu2017,podhora2017,zugenmaier2018,wang2018}. The demonstrated suitability of warm atomic vapors for generation of biphotons with spectral bandwidths on the order of hundreds of MHz together with the recent demonstrations of high-bandwidth optical memories utilizing the same experimental platform~\cite{guo2019,reim2011,wolters2017,kaczmarek2018} promise a viable approach for scalable, high rate, and high bandwidth distribution of quantum correlations.

We report on the enhancement of a generation of nonclassical photon pairs in warm $^{87}$Rb vapor in the process of spontaneous four-wave mixing (SFWM) implemented in a double $\Lambda$-type atomic energy level scheme. We observe unprecedentedly high nonclassical correlations, which result from the combination of several important mechanisms for reducing the thermal atomic effects, including the combination of particular excitation regime close to the vapor cell viewport and adaptation of optical pumping technique with paraffin coated cell for reducing the Raman noise in the anti-Stokes field mode. Importantly, the excitation close to the optical viewport seems to be crucial for the generation of several 100~MHz bandwidth fields in the fast resonant anti-Stokes Raman process, which would otherwise result in resonant absorption and uncorrelated photon noise. We analyse the residual noise contributions by precise characterization of the observed two-photon coupling efficiency and identify possible ways for further improvements.

\section{Generation of biphotons using SFWM in warm atomic vapors}
\label{sec:experiment}
\begin{figure*}[!t]
\includegraphics[width=180mm]{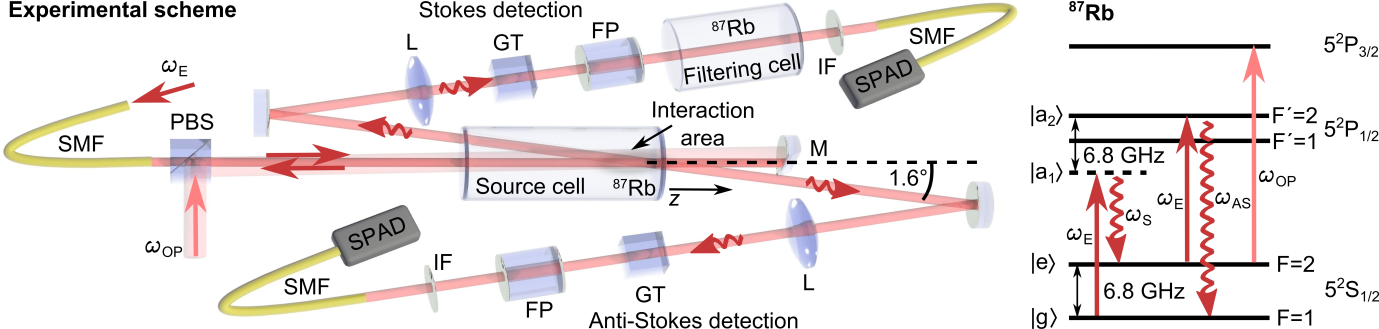}
\caption{Illustration of the employed experimental scheme. The relevant energy level scheme of $^{87}$Rb incorporates the excitation of the SFWM process following the $|g\rangle\rightarrow|a_1\rangle\rightarrow|e\rangle\rightarrow|a_2\rangle\rightarrow|g\rangle$ transition, which includes the spontaneous emission of Stokes and anti-Stokes photons with the relative energy difference of 13.6~GHz. The simplified experimental scheme depicts the SFWM excitation setup, optical pumping geometry, and optical filtering and detection setups. The optical pumping (OP) directly overlaps with the interaction area to suppress the contribution of atoms which were depolarized upon the interaction with the optical viewport. The polarization of the excitation beam (E) is set to linear by the polarization beam splitter (PBS). The scattered photon pairs collected under a small angle with respect to excitation laser beams are collected using the pair of identical lenses $L$ and polarization filtered by Glan-Thompson polarizers (GT). The transmitted linear polarization is perpendicular to the polarization of the excitation laser. The frequency filtering by a set of Fabry-P\'erot (FP) resonators, interference frequency filters (IF) and $^{87}$Rb vapor cell in the Stokes detection, effectively remove the contribution of various noise contributions. The single-mode fiber (SMF) coupled single-photon avalanche photodiodes (SPAD) are optimized for detection of the same spatial mode.}
 \label{fig:scheme}
\end{figure*}

The SFWM is implemented by excitation of the $^{87}$Rb vapor in double-$\Lambda$ type electronic level structure, see Fig.~\ref{fig:scheme} for the simplified experimental arrangement and relevant energy level scheme. Some details about the geometry, phase matching, and excitation parameters of the presented counter-propagating single laser frequency SFWM excitation can be found in~\cite{podhora2017}. The SFWM process is implemented in atomic vapor cell filled with $^{87}$Rb isotope and uses single laser locked to the resonance with $5{\rm S}_{1/2}({\rm F}=2)\leftrightarrow 5{\rm P}_{1/2}({\rm F}=2)$ for off-resonant excitation of the Stokes transition $|g\rangle = 5{\rm S}_{1/2}({\rm F}=2)\rightarrow |a_1\rangle\rightarrow |e\rangle= 5\mathrm{S}_{1/2}({\rm F}=2)$, from where the electron is quickly excited to $|a_2\rangle = 5\mathrm{P}_{1/2}({\rm F=2})$ followed by the emission of anti-Stokes photon at $|a_2\rangle\rightarrow|g\rangle$ transition. The virtual level $|a_1\rangle$ is approximately $6.8$ GHz detuned from the $|a_2\rangle$ state. The important distinctions with respect to our previous realization of single excitation laser SFWM scheme correspond to the adaptation of the auxiliary optical pumping on the 5S$_{1/2}(\rm F=2)\leftrightarrow 5{\rm P}_{3/2}$ transition, paraffin coated 7.5~cm long vapor cell, modified frequency filtering scheme, and implementation of the precise setting of the interaction area position with respect to the atomic cell output viewport.

SFWM in warm atomic ensembles  utilizes coherent collective spin superposition of a large number of atoms. The coherence of this process is limited by atomic thermal motion and thus the emission of anti-Stokes photon should happen on short timescales comparable to the escape time of atom from the area given by its thermal de Broglie wavelength and velocity corresponding to the photon recoil energy~\cite{mitchell2000dynamics}. This yields the typical small angle scattering configurations employed so far among all biphoton sources in warm atomic vapors~\cite{chen2008,willis2010,ding2012,shu2016,lee2016,zhu2017,podhora2017}. We set up the angle between excitation laser beams and observation directions to $\theta=1.6\pm 0.1$\,$^\circ$. The purity of biphotons generated in the double-$\lambda$ energy atomic level structure can be strongly deteriorated by Raman noise in the anti-Stokes mode coming from thermal drifts of atoms with population in $|e\rangle$ state into the interaction area. This noise can be efficiently suppressed by the employment of atomic polarization preserving coatings of vapor cells together with auxiliary optical pumping beam exciting atoms close to the interaction area~\cite{shu2016}, or using doughnut beams~\cite{zhu2017}. We employ the paraffin coating on the inner surfaces of the quartz glass vapor cell, with two optical quality viewports anti-reflection coated for both 780~nm and 795~nm. The auxiliary optical pumping is implemented by overlapping a 780~nm laser beam resonant with the 5S$_{1/2}({\rm F} = 2)\leftrightarrow 5{\rm P}_{3/2}$ manifold with 43~mW of optical power and approximately 3~mm width (FWHM) at the position of the interaction region. Here and in the rest of this work, the effective interaction region is defined as an overlap of observation and excitation SFWM optical modes. The optical pumping beam is propagating approximately collinearly with respect to the 795~nm excitation beam. Crucially, the resulting coverage of large area around the interaction region results in efficient optical pumping before the atoms enter the interaction area. At the same time, the optical intensity of 780~nm beam being more than two orders of magnitude lower than the intensity of the 795~nm excitation laser effectively avoids the contribution of Raman noise to Stokes detection mode. In contrast to previous demonstrations of the optical pumping effect on bi-photon generation~\cite{shu2016,zhu2017}, the spatial optimization of the presented source results in the minimum Raman noise for the optical pumping directly overlapping with the interaction region. This is a consequence of the proximity of the interaction region to the output optical viewport, which is kept free from polarization preserving coating in order to avoid optical abberations and strong scattering of the laser beam.

Optical modes of the Scattered Stokes and anti-Stokes fields are defined using the polarization, frequency and spatial filtering in two exactly opposite detection directions. The polarizations of detected photons are set by the pair of Glan-Thompson (GT) polarizers with the transmitted polarization oriented perpendicularly with respect to the polarization of the excitation laser. The frequency filtering by a pair of Fabry-P\'erot (FP) etalons with transmission peaks centered at Stokes and anti-Stokes frequencies is used mainly for the suppression of detection of light resonant with the $|a_2\rangle\rightarrow|e\rangle$ transition. Both etalons have the transmission linewidth $\sim 0.8$~GHz (FWHM), free spectral range of about 30~GHz, and the peak transmissivity of 71~\% and 62~\% for anti-Stokes and Stokes etalons, respectively. Importantly, the frequency filtering takes place also in the $^{87}$Rb cell itself. As the interaction region is chosen to be close to the cell window in order to prevent the anti-Stokes photons from being reabsorbed, Stokes field propagates through almost whole cell length resulting in an efficient filtering of resonant laser beam scattering. We note, that this configuration effectively corresponds to using short SFWM cell combined with additional filtering cell in the Stokes emission direction. In order to effectively enhance the optical depth of this filter and make it partially independent of the atomic density in the generation cell, we use additional $^{87}$Rb cell heated to 80$^\circ$C is placed in the Stokes detection mode behind the optical cavity. Any stray light from the pump laser is very efficiently suppressed using the bandpass interference filters (IF) mounted directly on the fiber coupling assembly and with power extinction $10^5$ at 780~nm. The spatial mode of biphotons is defined by a combination of two collimating lenses $L$ with equal focal length of 30~cm and coupling to single-mode optical fibers. The corresponding Gaussian mode waist at the interaction region is $290~\pm 20~\mu$m. The single-mode fibers in both detection channels are connected to single-photon avalanche photodiodes (SPAD). The measured total detection efficiency at the peak transmission frequency of Stokes and anti-Stokes photons entering detection spatial modes is $\sim 21$\,\% and $\sim 23$\,\%, respectively, including the specified efficiencies of APDs. The intensity of the excitation beam in the interaction region is further enhanced by its focusing corresponding to Gaussian beam waist width $290~\pm 20~ \mu$m at the position of retro-reflecting mirror (M). The distance of the interaction area from the mirror is 8~cm and we did not observe any clear degradation of the biphoton correlations when compared to the collimated excitation beam of a similar diameter. However, the observed two-photon correlations are very sensitive to the relative position of the beam focus and the mirror (M), with the optimal waist position corresponding precisely to the position of the mirror and thus to optimal phase matching conditions.

We note that in contrast to employment of ladder energy level schemes~\cite{ding2012,willis2010,lee2016} or double-$\Lambda$ schemes including Stokes and anti-Stokes transitions with large energy difference on the order of energy of fine splitting~\cite{chen2008,shu2016,zhu2017}, the employed SFWM level scheme comprising solely hyperfine levels of the 5S$_{1/2}(F = 2)\leftrightarrow 5P_{3/2}$ transition and counter-propagating excitation laser beam guarantees the optimality of phase-matching for detection modes with exactly opposite directions~\cite{kolchin2006,podhora2017}. This configuration thus substantially simplifies the whole experimental alignment as it in principle doesn't require any auxiliary seed beams. The case of single excitation laser further enhances the technological simplicity of the scheme at the expense of the inevitable small residual phase mismatch of the SFWM process corresponding to the frequency difference of Stokes and anti-Stokes fields $\Delta\nu_{\rm S,AS}\sim 13.6$ GHz. This imposes the lower limit on the emission angle between laser excitation and detection directions, given by the spatial length $d=c/(2 \Delta\nu_{\rm S,AS})\sim 11$\,mm corresponding to the phase mismatch of $\pi$. Therefore, the excitation and detection modes should overlap only within the distance shorter than $d$ for optimal biphoton generation. The excitation beam diameter of $2w_l=0.35\pm0.02$~mm together with the detection mode diameter of $\sim 0.1$ mm yield optimal observation angle of $2.3^\circ$. Experimentally optimized angle of $\theta=1.6\pm0.1^\circ$ corresponding to 16~mm of interaction area length is in approximate agreement with the theoretical value when taking into an account the shortening of the interaction region due to the proximity of the optical viewport, which was estimated at maximum correlation measurements to cut away $6\pm 2$~mm.

\section{Measurements and characterizations of nonclassical correlations}
\label{sec:source}

We analyze the observations of generated nonclassical light by evaluation of violation of Cauchy-Schwartz inequality together with measurements of the generated bi-photon rates, spectral bandwidths, and overall heralding efficiency. The measurement of photodetection events on pairs of single photon detectors in two spatial modes $i,j$  can be evaluated in the form of fsecond order correlation function $g^{(2)}_{\rm i,j}(\tau)=\langle \hat{a}_{\rm i}^\dagger (t_{\rm i})a_{\rm j}^\dagger (t_{\rm i}+\tau) \hat{a}_{\rm j}(t_{\rm i}+\tau)\hat{a}_{\rm i}(t_{\rm i})\rangle/(\langle \hat{a}_{\rm i}^\dagger \hat{a}_{\rm i}\rangle\langle \hat{a}_{\rm j}^\dagger \hat{a}_{\rm j}\rangle)$, where $\hat{a}^\dagger$ and $\hat{a}$ are creation and annihilation operators, and $\tau$ is the time delay between the successive photon detection events on the two detectors. All classical light states must satisfy the Cauchy-Schwarz (C-S) inequality corresponding to $(g^{(2)}_{S,AS}(\tau))^2\geq g^{(2)}_{S,S}(0)g^{(2)}_{AS,AS}(0)$. Although its violation is a bare witness of nonclassical correlations~\cite{loudon2000quantum}, from the application perspective, the high degree of violation of C-S inequality favors the successful preservation of nonclassical correlations upon the interaction of correlated photons with lossy or noisy environment~\cite{eisaman2005,rielander2014quantum,jin2015telecom,distante2017storing,zugenmaier2018,seri2019quantum,tsai2019quantum}.

\begin{figure*}[!t]
 \begin{minipage}{0.45\textwidth}
\hspace{-4mm}\includegraphics[width=73mm]{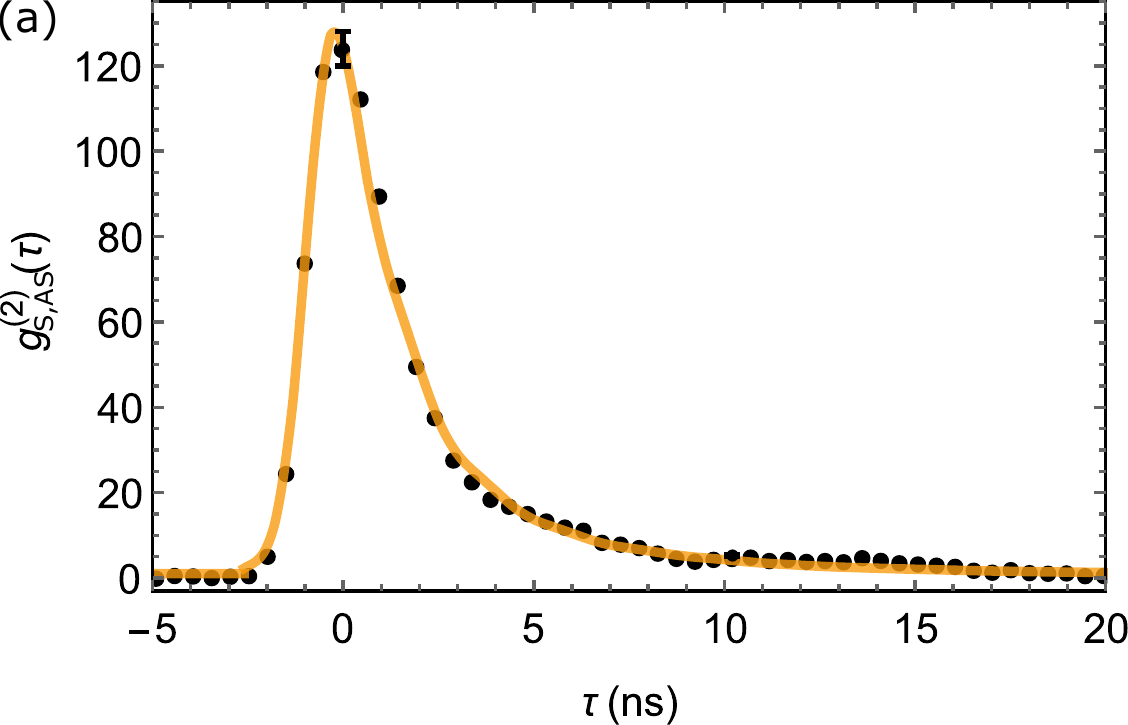}
 \end{minipage} \hfill
 \begin{minipage}{0.45\textwidth}\vspace{1mm}
\hspace{-13mm}\includegraphics[width=81mm]{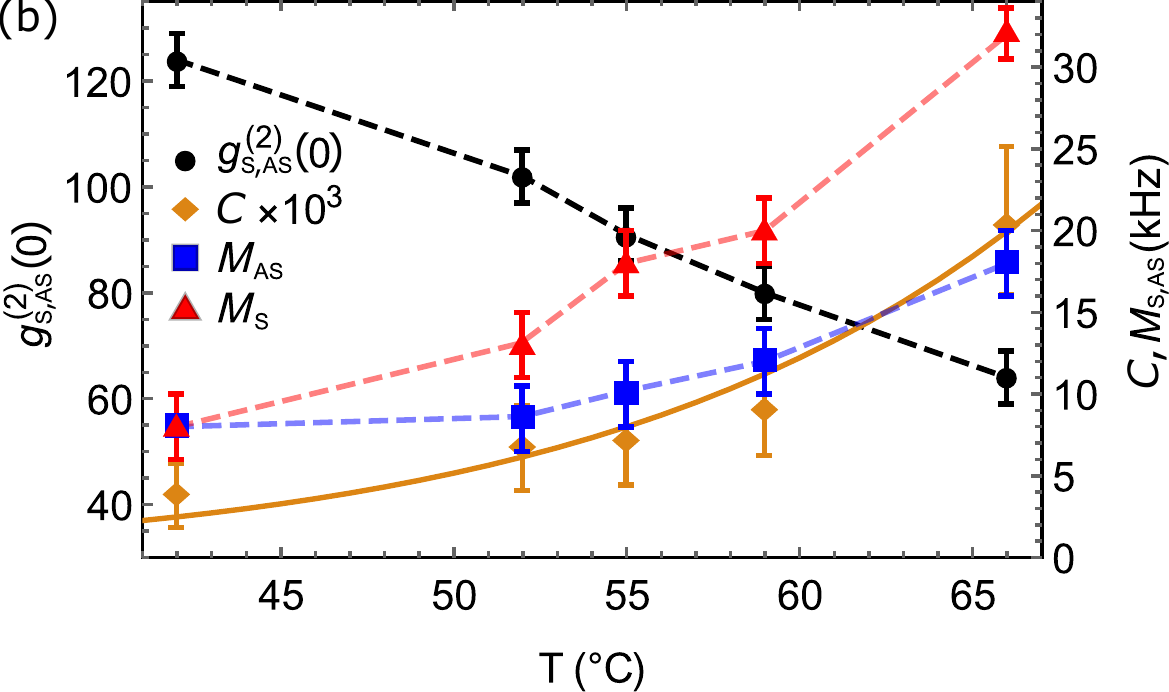}
 \label{fig:temp}
 \end{minipage}
\caption{a) The representative measurement of the second order correlation function $g^{(2)}_{\rm S,AS}(\tau)$ between Stokes and anti-Stokes photons. The solid curve corresponds to a theoretical model, see text for details. The inset shows the evaluated maximal values of $g^{(2)}_{\rm S,AS}(0)$ for different temperatures of the source $^{87}$Rb cell. b) The measured dependence of detected Stokes (blue squares), anti-Stokes (red triangles), and coincidence count rates (brown circles) evaluated for various temperatures. The solid curve corresponds to a theoretical model. The coincidence time window was chosen as the full width at half maximum value (FWHM) of each corresponding temporal profile of $g^{(2)}_{S,AS}(\tau)$. The dashed lines merely connect the measured points to guide the eye. Error bars correspond to a single standard deviation.}
\label{fig:max}
\end{figure*}

The Fig.~\ref{fig:max}~a) shows the $g^{(2)}_{S,AS}(\tau)$ measured for the excitation beam tuned in resonance with the $|e\rangle\rightarrow|a_2\rangle$ transition, excitation optical power of 30~mW, and source cell heated to 42~$^\circ$C. The maximum of the normalized correlation is $g^{(2)}_{S,AS}(0)=124\pm4$. The autocorrelation functions maxima were measured as $g^{(2)}_{AS,AS}(0)=1.74\pm0.02$ and $g^{(2)}_{S,S}(0)=1.60\pm0.02$ for coincidence window of $\tau=486$~ps. This corresponds to violation factor of C-S inequality $F=[g^{(2)}_{S,AS}(\tau)]^2/(g^{(2)}_{S,S}(0)g^{(2)}_{AS,AS}(0))=(5.6\pm0.1)\times 10^3$. To our best knowledge, this corresponds the highest value of both $g^{(2)}_{S,AS}(0)$ and C-S violation observed in discrete correlations of photons generated in warm atomic source in $\Lambda$-type energy level scheme, with $g^{(2)}_{S,AS}(0)$ increased by more than factor of two from the previous record~\cite{zhu2017} and becoming comparable to the SFWM nonclassical light sources implemented with cold atomic ensembles in this type of atomic energy level scheme~\cite{du2008narrowband}. We note that with the assumption of measurement of correlations on pairs of single photons, the estimated of the corresponding degree of anti-bunching of anti-Stokes field heralded by Stokes detection gives $g^{(2)}_{\rm her}(0)=0.028\pm0.002$~\cite{lee2016,podhora2017}.
The measured temporal width of $\sim 2.7$~ns corresponds to the estimated spectral bandwidth of $\Delta\nu\sim 370$~MHz. The presented data were acquired in 15~minutes long measurement and were corrected for detector dark counts, which were separately measured as $B_{\rm S}=0.2\pm 0.1$~kHz and $B_{\rm AS}=2.2\pm 0.2$~kHz on Stokes and anti-Stokes detectors, respectively. The normalized correlation without any background correction corresponds to $g^{(2)}_{\rm S,AS}=97\pm1$ and $F=3390\pm70$. The measured count rates were $8\pm1$ kHz in both Stokes and anti-Stokes detection channels. The solid orange line in Fig.~\ref{fig:max}~a) corresponds to the evaluated theoretical model of SFWM in warm atoms~\cite{du2008narrowband,wen2006transverse,wen2007biphoton,shu2016}. The particular experimental arrangement of excitation near output optical viewport, single SFWM excitation laser and thus, far off-resonant Stokes generation and low atomic density, considerably simplify the overall SFWM description resulting in the possibility of effectively neglecting the first order susceptibilities. At the same time, the near-collinear excitation allows for considering a bare 1D-Doppler broadening of spectral lines. The presented theoretical curve uses no free fitting parameters besides its vertical scaling. All model parameters including the pump power, atomic density and detuning of the excitation laser from the atomic resonance were estimated in independent measurements and the resulting function $g^{(2)}_{S,AS}(t)$ was convolved with a temporal filter corresponding to measured Lorentzian spectral profile of employed Fabry-P\'erot filters with measured width $\Gamma= 896\pm 5$ MHz.

To better understand the properties of the presented SFWM photon source, we devise a simple theoretical model. It considers a true photon pair production rate at the position of the atomic cell emitted in the given detection modes $P$, independently estimated overall detection efficiencies $\eta_{\rm S}$ and $\eta_{\rm AS}$, measured count rates $M_{\rm S}$ and $M_{\rm AS}$, and individual channel noise contributions $N_{\rm S}$ and $N_{\rm AS}$, corresponding to detection in Stokes and anti-Stokes channels, respectively,
\begin{eqnarray}  \label{eq:model1}
M_{\rm S}= & \eta_{\rm S}(P+N_{\rm S})+B_{\rm S}, \\
M_{\rm AS}= & \eta_{\rm AS}(P+N_{\rm AS})+B_{\rm AS}. \nonumber
\end{eqnarray}
Here, we assume that the detected noise originates from the source cell and thus experiences the same losses as signal photons, up to the limited background counts $B_{\rm S}, B_{\rm AS}$ which include also detector dark counts. The measured number of coincidences $C$ within the time window $\Delta t$ at zero time delay $\tau=0$ can be evaluated by summing the SFWM generated biphotons with random coincidences corresponding to noise $C  = C_{\rm sg} + C_{\rm ns}$, where
\begin{eqnarray}  \label{eq:model2}
C_{\rm sg}= & \eta_{\rm S} \eta_{\rm AS} P, \\
C_{\rm ns}= & (\eta_{\rm S}\eta_{\rm AS} (P\cdot N_{\rm S}+P\cdot N_{\rm AS}+N_{\rm S} N_{\rm AS})+ \nonumber\\
   & +  \eta_{\rm S}(P+N_{\rm S})B_{\rm AS}+\eta_{\rm AS}(P+N_{\rm AS})B_{\rm S} +  B_{\rm S} B_{\rm AS})\Delta t. \nonumber
\end{eqnarray}
Using this simple model, estimation of the optical losses can allow for the determination of individual noise contributions $N_{\rm S}, N_{\rm AS}$, which can in turn reveal the portion of uncorrelated photon pairs at the source position and the two photon coupling efficiency. The complementary evaluation can employ the knowledge of the correlation function maximum $g^{(2)}_{S,AS}(0)$, which corresponds to the ratio of signal to background coincidences for coincidence window $\Delta t$ much smaller than the temporal width of measured $g^{(2)}_{S,AS}(\tau)$,
\begin{equation} \label{eq:g2}
g^{(2)}_{S,AS}(0)\approx 1 + C/C_{\rm ns}.
\end{equation}
Importantly, the employed SFWM setup and energy level scheme necessarily bind the presence of competing SFWM processes with the same level structure but different geometry, which result in minimal amount of noise for the given scheme and impose a fundamental limit on the achievable $g^{(2)}_{S,AS}(0)$. In the excitation configuration corresponding to of two counter-propagating excitation frequency degenerate beams with frequency selective detection of Stokes and anti-Stokes fields, this amounts for in total six competing SFWM processes, all with the near equal probability.
In addition to this noise, the finite coherence of the $|g\rangle\leftrightarrow|e\rangle$ transition and atomic thermal motion~\cite{mitchell2000dynamics} can limit the spatially correlated emission of anti-Stokes photons which, together with imperfect two-photon coupling and spatial mode matching, results in the rest of the observed noise signal. Evaluation of the equations~\ref{eq:model1}-\ref{eq:model2} for the values of noise and signal contributions from the measurement presented in the Fig.~\ref{fig:max}-a) corresponding to $M_\mathrm{S}=8\pm1.5~\mathrm{kHz}$, $M_\mathrm{AS}=10\pm2~\mathrm{kHz}$, $\eta_{\rm S}=0.21\pm0.02$, $\eta_{\rm AS}=0.22\pm0.01$, $C=18.5\pm0.1~{\rm Hz}$, $\tau=2.916~{\rm ns}$ results in $P=383\pm3~{\rm Hz}$ and $N_{\rm S}=35\pm5~{\rm kHz},N_{\rm AS}=33\pm4~{\rm kHz}$. Here, the competing SFWM processes corresponding to observable Raman noise amount for $N_{\rm S}=31\pm5~{\rm kHz}$ and $N_{\rm AS}=25\pm3~{\rm kHz}$.

The detected single and coincidence rates for different temperatures of the atomic source cell are shown in the Fig.~\ref{fig:max}~b).
We observe relatively low detectable coincidence rates and monotonically increasing dependence of number of atoms $n$ as a function of the cell temperature $T$ in agreement with the evaluated density of Rb atoms for given $T$~\cite{steck2001rubidium}. This, together with the opposite dependence of the $g^{(2)}_{\rm S,AS}(0)$ compromises particular working point which should be optimized for individual applications. As a viable solution, we envisage the possibility of increasing the biphoton detection rate by enhancement of the interaction area volume using radially increased excitation and detection mode diameters and proportionally increased excitation laser power, which has been in our case limited to approximately 40~mW.
The black points in Fig.~\ref{fig:max}~b) show the measured dependence of the cross-correlation maximum $g^{(2)}_{S,AS}(0)$ on temperature of the source cell. 
Assuming the ratio of detectable photon pairs to uncorrelated noise photons does not change with number of atoms for sufficiently high atomic densities, the observable linear decrease of the $g^{(2)}_{\rm S,AS}(0)$ is deducible from the Eq.~\ref{eq:model1} and~\ref{eq:model2}. A linear dependence of generated photon rates on number of contributing atoms $n$ for both photon pairs and noise photons $P(n), N_{\rm S}(n), N_{\rm AS}(n) \sim n$ results is in a quadratic dependence of noise coincidences $C_{\rm ns}$ and a linear dependence of the signal pairs $C_{\rm sg}$ on the atomic density.

\begin{figure*}[!t]
\begin{minipage}{0.45\textwidth}
\hspace{-4mm}\includegraphics[width=80mm]{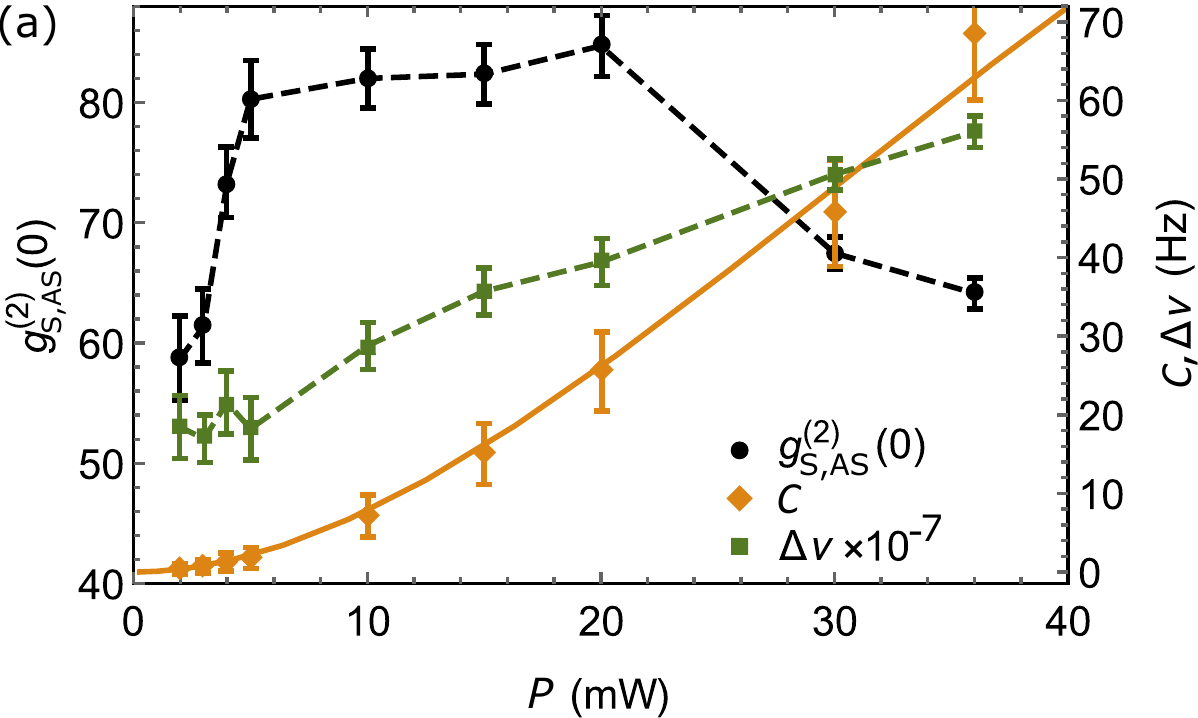}
\label{fig:power}
\end{minipage} \hfill
\begin{minipage}{0.45\textwidth}\vspace{0mm}
\hspace{-8mm}\includegraphics[width=78mm]{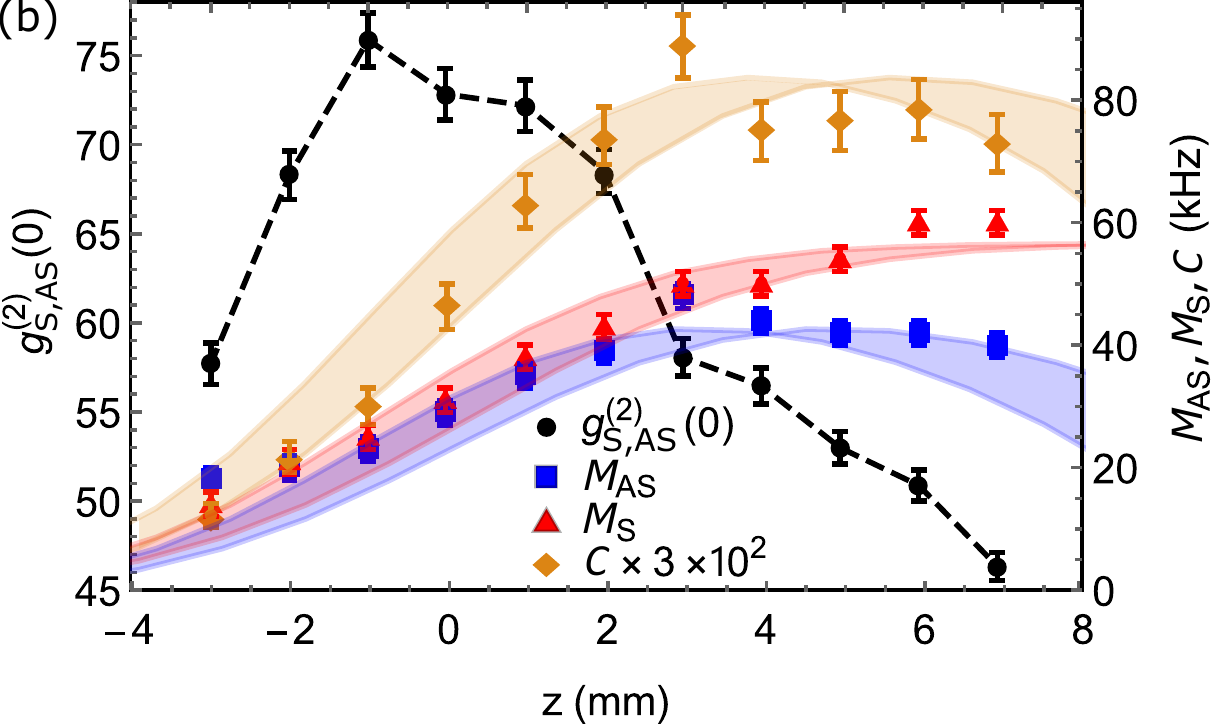}
\label{fig:displacement}
\end{minipage}
\caption{a) shows the estimated single standard deviation of the emitted photons $\delta\nu$ together with the measured $g^{(2)}_{\rm S,AS}(0)$ values and corresponding coincidence rates for the accessible scale of 795~nm excitation powers $P$. The measured effect of atomic cell displacement along the direction $z$ on the detected photon rates and $g^{(2)}_{\rm S,AS}(0)$ is shown in graph b). Here, the positive cell displacement $z$ corresponds to moving the cell towards the mirror $M$. The zero displacement $z=0$ corresponds to the position of the cell output viewport at the center of the interaction area found by the best fit of the theoretical model for the Stokes photon rates to the data, see the text for details. The dashed lines connect the points to guide the eye and the solid lines corresponds to the theoretical models. The uncertainty in the measurement of the angle~$\theta$ between excitation and emission optical modes results in the simulated uncertainties of the detected photon rates shown as filled areas, with borderlines corresponding to a single standard deviation.}
\end{figure*}

We further presents the measurements of phenomenologically exquisite characteristics of the realized SFWM photon source. These especially include its dependence on the excitation beam power and on the relative position of the interaction area and atomic cell viewport. When compared to the excitation schemes with an independent pump-coupling laser beams~\cite{chen2008,shu2016,zhu2017}, the presented approach employing a single 795~nm excitation laser results in the contribution of different number of competing SFWM processes. In addition, the spatial overlap of the excitation and optical pumping beams can limit the SFWM process efficiency for small 795~nm intensities.
The Fig.~\ref{fig:power}-a) depicts the measured $g^{(2)}_{\rm S,AS}(0)$ as the function of the 795~nm laser power. The atomic cell temperature in these measurements corresponds to 59~$^\circ$C and the excitation laser is locked to the resonance with the $|e\rangle\leftrightarrow|a_2\rangle$ transition. The observable initial steep enhancement of the correlation can be mostly attributed to the suppression of the additional Raman noise corresponding to depopulation of the $|e\rangle$ levels by the 780~nm beam and to overcoming the decoherence corresponding to short thermal de Broglie wavelength of warm atoms~\cite{mitchell2000dynamics}. For the excitation beam power of $P=2$~mW, the intensity of the 795~nm beam at the observation region is only about a factor of 10~higher than that of optical pumping beam. For higher excitation beam powers these effects become negligible, however,  the linear dependence of $N_{\rm S}, N_{\rm AS}$ and $C_{\rm sg}$ on the excitation power $P$ results in a linear decrease of the two-photon correlation amplitude according to the~Eq.~\ref{eq:g2}. The spectral widths of the emitted two-photon wavepackets evaluated as the Fourier-limited functions corresponding to measured temporal profiles $g^{(2)}_{\rm S,AS}(\tau)$ demonstrate the tunability of the output frequency bandwidths from approximately 180 to 560~MHz, while preserving very high nonclassical correlations beyond $g^{(2)}_{\rm S,AS}(0)\sim 60$.

Fig.~\ref{fig:power}-b) depicts the source parameters for different relative positions of the interaction area and the atomic cell output viewport along the direction $\vec{z}$. Here, the positive displacement corresponds to shif the atomic cell towards the retro-reflecting mirror and thus increasing the overlap of atomic vapor with the interaction area, see the spatial configuration the Fig.~\ref{fig:scheme}. The shift of the relative position of the interaction area outside of the atomic cell ($z<0$) results in expectable decrease of all measured count rates, as physically fewer atoms can be excited. The corresponding linear decrease of the $g^{(2)}_{\rm S,AS}(0)$ can be explained by considering the effect of absolute number of contributing atoms, in analogy with the case of varying atomic density presented in~Fig.~\ref{fig:max}-b).
At the same time, relative position $z<0$ also leads to higher relative contribution of atoms experiencing small excitation beam intensity. The longer average excitation times on the $|e\rangle\rightarrow|a_2\rangle$ transition result in an increased probability of depopulation of $|e\rangle$ state by 780~nm beam and in partial decoherence of the collective spin excitation. In addition, the contribution of atoms positioned very close to the cell viewport increases the probability of $|e\rangle\rightarrow|a_2\rangle\rightarrow|g\rangle$ Raman noise due to a high probability of interaction of atoms with the uncoated quartz window. Phenomenologically interesting behaviour can be observed also for the shift of the interaction area to positive displacement values, which results in observable absorption of anti-Stokes photons on their way towards the detector. For large positive displacements the Stokes detection rate tends towards the saturation, because almost full interaction volume can contribute to their production. The anti-Stokes photons experience increased absorption probability, which results in the corresponding reduction of observable $g^{(2)}_{\rm S,AS}(0)$ due to a higher relative contribution of noise. The measured singles and coincidence rates are reproduced by the equations~\ref{eq:model1}-\ref{eq:model2}, where the effect of position dependence has been taken into an account by considering the number of contributing atoms within the effective interaction volume. The interaction volume has been evaluated as a spatial overlap of the bulk cell with the  excitation and observation gaussian spatial modes. The best fit of the dependence for the Stokes photons has been used for the estimation of zero displacement $z=0$, which corresponds to the position of the center of interaction area at the inner face of the cell viewport. A measured optical depth of the source cell on the $|g\rangle\rightarrow|a_2\rangle$ transition in the presence of the optical pumping beam $OD=3.4\pm 1.4$ was used for including the effect of absorption of anti-Stokes photons, which results in a good agreement of the evaluated anti-Stokes photon rate model shown as a blue area in the Fig.~\ref{fig:power}-b) with the measured data, with no free fitting parameters. The borders of the marked areas correspond to the uncertainty of single standard deviation of the measured scattering angle $\theta=1.6\pm 0.1\,^\circ$.
The observed temporal width of the cross-correlation $g^{(2)}_{\rm S,AS}(\tau)$ was steadily increasing on the scale of presented position shifts with corresponding estimated spectral widths ranging from $\delta\nu=340\pm 10$~MHz to $\delta\nu=630\pm 20$~MHz for $z=-3$ mm and $z=7$ mm, respectively. We attribute this spectral behaviour to the change of the average excitation intensity for atoms present in the interaction area, which effectively increases with the shift of the interaction area towards the atomic cell.
While the presented large bandwidth regime can be clearly beneficial for a high-speed quantum communication and interaction with available broadband quantum memories based on the same experimental platform~\cite{guo2019,reim2011,wolters2017,kaczmarek2018}, the correlation peak $g^{(2)}_{\rm S,AS} (0)$ is naturally enhanced by reaching the corresponding short timescale of the SFWM process and overcoming of decoherence mechanisms with a characteristic time given by the atomic thermal de Broglie wavelength and kinetic energy of photon recoils~\cite{mitchell2000dynamics}. In order to facilitate comparison of the achieved results with the state of the art experiments operating in a sub-natural linewidth regime, we have realized a measurements for different detunings $\Delta$ of the excitation laser from the $|e\rangle\rightarrow|a_2\rangle$ transition. We note, that this parameter has been chosen due to the relative simplicity and traceability of its change and more optimal procedures corresponding to more complex parameter combinations for tuning between the two bandwidth regimes could be suggested. The observed value of $g^(2)_{S,AS} (0)=29\pm 2$ at the achieved frequency bandwidth of $\Delta\nu=66\pm 5$~MHz for a detuning $\Delta \sim +1$~GHz is comparable to the values presented in reference~\cite{shu2016} operating with similar optical pumping scheme. We would like to remind, that these experiments operate in a phenomenologically different regime by employing an EIT assisted SFWM, which is hard to extend to a several hundreds of MHz regime presented here.

\section{Conclusions}
The generation of nonclassically correlated photon pairs capable of efficient interaction with target atomic ensembles corresponds to a lively experimental goal in quantum optics community with a broad range of applications including the heralded generation of nonclassical light, the generation of photonic entanglement, or distribution of quantum information using atoms and photons. We have presented the source of photon pairs based on the excitation of the SFWM process in warm $^{87}$Rb vapor employing simultaneously two crucial mechanisms for the enhancement of the nonclassical photon correlations between Stokes and anti-Stokes fields. First, the proximity of the interaction area close to the optical viewport guarantees the low absorption of the AS photons upon leaving the atomic medium in a broad spectral range, which allows for a biphoton correlation times on the scale of several nanoseconds and reduction of the decoherence of the SFWM process due to photon recoils~\cite{podhora2017}. At the same time, the additional optical pumping limits the typical large contribution of Raman noise in the AS detection mode~\cite{shu2016}. The combination of these two mechanisms provides a feasibility of unprecedentedly high $g^{(2)}_{\rm S,AS}(0)$ regime in warm atom SFWM sources. Demonstrated nonclassical correlations of $g^{(2)}_{\rm S,AS}(0)=124\pm4$ violate the classical boundary given by the Cauchy-Schwarz inequality by a factor of $F=5600\pm100$ corresponding to the highest reported values for the warm atomic vapor source of biphotons utilizing the convenient $\lambda$-type energy level scheme. The value of the normalized two photon correlations has thus been improved by more than an order of magnitude compared to the previous realization with single laser excitation~\cite{podhora2017} and approximately by a factor of two with respect to the best reported values so far~\cite{zhu2017}. In addition, the scheme did not employ a hollow-beam for optical pumping, which is expected to further enhance the $g^{(2)}_{\rm S,AS}(0)$. The short temporal profiles of photon correlations corresponding to spectral bandwidths of up to 560~MHz promise a direct applicability of this source for realization of storage of nonclassical light in large-bandwidth quantum memories implemented in same experimental platforms~\cite{guo2019,reim2011,wolters2017,kaczmarek2018}. Such coherent storage would further allow for realizations of on-demand single photon sources and long distance quantum communication based on the feasible technology.
The observed dependence of the generated photon rates on the relative position of the interaction region and the output viewport of the atomic cell demonstrates the relevance of the presented approach for high bandwidth applications. The proximity of the interaction area to the cell viewport has proved to be crucial for the achievement of high photon correlations between Stokes and anti-Stokes fields with the observable significant anti-Stokes attenuation and $g^{(2)}_{\rm S,AS}(0)$ suppression on the scale of millimeter position shifts. The alternative approaches for suppression of the anti-Stokes absorption could be based on the off-resonant excitation of the 5S$_{1/2}({\rm F=2})\leftrightarrow{\rm 5P}_{1/2}({\rm F=2})$ transition, or on employing the electromagnetically induced transparency~\cite{shu2016}, with the expectable compromise in the spectral bandwidths due to the practically demanding achievement of transmission windows on the order of hundreds of MHz~\cite{wolters2017}. The presented approach allows for a substantial simplification of the theoretical description of the biphoton generation and promises the feasibility of generation of intensity squeezing in new excitation configurations and spectral bandwidth regimes~\cite{mccormick2007strong,turnbull2013}. The presented high nonclassical correlations should also allow for the generation of two-mode states entangled in various degrees of freedom with high fidelity. The entanglement generation based on the SFWM in warm atomic vapors has been recently demonstrated in several notable works employing interferometric excitation of a cascade photon emission~\cite{Park:19}\cite{Wang:19}\cite{Moon2019}\cite{Wang:20}, however, the feasibility with the convenient double-lambda energy level scheme remains to be proven.

\section*{Acknowledgments}

This work has been supported by the grant No. GA19-14988S of the Czech Science Foundation and Palacky University IGA-PrF-2019-010.

\bibliography{biphoton_source}

\end{document}